\documentclass[12pt]{article}

\usepackage[titletoc,title]{appendix}
\usepackage[margin=1in]{geometry}
\usepackage{array}
\usepackage{color}
\usepackage{amsmath,bm}
\usepackage{bbm}
\usepackage{amssymb}
\usepackage{amsthm}
\usepackage{amsfonts}
\usepackage{enumerate}
\usepackage[toc]{glossaries}
\usepackage{hyperref}
\usepackage{listings}
\usepackage{natbib}
\usepackage{url}
\usepackage{graphicx}
\usepackage{graphics}
\usepackage{subcaption}
\usepackage{fancyhdr}
\usepackage{rotating}
\usepackage{multirow}
\usepackage[usenames,dvipsnames]{xcolor}
\usepackage{sectsty}
\usepackage{titlesec}
\usepackage{booktabs}
\usepackage{datetime}
\usepackage{tikz}
\usetikzlibrary{arrows,automata,positioning,chains,fit,shapes,calc,patterns}
\usepackage{type1cm}
\usepackage{lettrine}
\usepackage{comment}
\usepackage{morefloats}
\usepackage{longtable}
\usepackage{psfrag}
\usepackage{changepage}
\usepackage{xstring}
\usepackage[ampersand]{}
\usepackage{mathtools}
\usepackage{multicol}
\usepackage{setspace}
\usepackage{censor}
\usepackage{tabularx}
\usepackage[space]{grffile}
\usepackage{blkarray, bigstrut} %
\usepackage[normalem]{ulem}
\usepackage[utf8]{inputenc}
\useunder{\uline}{\ul}{}

\settimeformat{ampmtime}
\mmddyyyydate

\usepackage{multirow,bigdelim}

\IfFileExists{/Users/jfogel/Networks/bibtex/bibtex.bib}{

}{

}

\definecolor{linkcolor}{rgb}{0,0,0.50}
\setcitestyle{authoryear, round, semicolon, aysep={},yysep={;},notesep={,}}
\hypersetup{
	bookmarks=true,                 % show bookmarks bar?
	unicode=false,                  % non-Latin characters in bookmarks
	pdftoolbar=true,                % show toolbar?
	pdfmenubar=true,                % show menu?
	pdffitwindow=true,              % window fit to page when opened
	pdfstartview={FitH},            % fits the width of the page to the window
	pdftitle={},                    % title
	pdfauthor={BPS team},           % author
	pdfnewwindow=true,              % links in new window
	colorlinks=true,                % false: boxed links; true: colored links
	linkcolor=black,                 % color of internal links
	citecolor=black,                % color of links to bibliography
	filecolor=magenta,              % color of file links
	urlcolor=blue,                  % color of external links
	breaklinks=true
}

\IfFileExists{/Users/jfogel/Networks/bibtex/bibtex.bib}{
	}{
	}

\setcitestyle{authoryear, round, semicolon, aysep={},yysep={;},notesep={,}}

\definecolor{gold}{rgb}{0.85,.66,0}
\definecolor{blue}{rgb}{0,0,1}

\def\bs{\begin{sideways}}
	\def\es{\end{sideways}}

\usepackage[draft]{fixme}
\fxsetup{layout=footnote, marginclue}

\usepackage{footmisc}

\theoremstyle{definition}
\theoremstyle{plain}

\usepackage{mathrsfs} % enable people to use \mathscr{A}

\makeglossaries

\title{Advancing Distribution Decomposition Methods Beyond Common Supports: Applications to Racial Wealth Disparities\footnote{University of Michigan, bmodene@umich.edu. I thank Mel Stephens, Florian Gunsilius, Gongju Xu, Abigail Jacobs for advice and guidance throughout this project. I also thank Jamie Fogel for helpful comments and discussions.}}

\author{Bernardo Modenesi}
\date{}
\begin{document}
	
	\maketitle

	\begin{abstract}
		I generalize state-of-the-art approaches that decompose differences in the distribution of a variable of interest between two groups into a portion explained by covariates and a residual portion. The method that I propose relaxes the overlapping supports assumption, allowing the groups being compared to not necessarily share exactly the same covariate support. I illustrate my method revisiting the black-white wealth gap in the U.S. as a function of labor income and other variables. Traditionally used decomposition methods would trim (or assign zero weight to) observations that lie outside the common covariate support region. On the other hand, by allowing all observations to contribute to the existing wealth gap, I find that otherwise trimmed observations contribute from $3\%$ to $19\%$ to the overall wealth gap, at different portions of the wealth distribution.
	\end{abstract}
	\clearpage

	%--------------------------
	
	\clearpage
	
	\onehalfspacing
%%%%%%%%%%%%%%%%%%%%%%%%%%%%%%%%%%%%%%%%%%%%%%%%%%%%%%%%%%%%%%%%%%%%%
\section{Introduction}

Decomposition exercises have been developed and improved in the Economics literature for decades, dating to the seminal papers of \citet{oaxaca1973} and \citet{blinder1973}. These tools decompose the difference between two groups with respect to a variable of interest $Y$ into a portion due to differences in observable group characteristics $X$ and another residual portion. If we consider $Y$ to be wealth, $X$ to be average lifetime labor income and the groups to be defined by blacks and whites, we would be able to compute the contribution of labor income in explaining the racial wealth gap with decomposition methods. 

A central part of decompositions involves estimating a counterfactual quantity which is what values of $Y$ would a group have if it had the $X$ distribution of the other group; e.g. what would be the white's wealth if they had the black's distribution of average lifetime labor income. In order to build this counterfactual, decomposition methods find blacks who are similar to each white with respect to all observable characteristics other than race and use these blacks' wealth and a model to predict it. An issue with this approach is that it is not always possible to find blacks who are similar to whites with regards to $X$. For instance, the range of the average lifetime labor income distribution was not the same for these races in the 1980's and beginning of 1990's. There is a nontrivial set of blacks exhibiting average income of nearly zero, with no white households earning that low; and with a considerable amount of whites at the top of the lifetime earnings distribution, unmatched by blacks. These observations are not informative for building counterfactual wealth for the other race. Current decomposition methods either trim these observations, or assign virtually zero weight to them in their estimation process, even though they may contribute significantly to the overall observed gap in $Y$.

In fact, the process of trimming observations that do not share similar $X$ values in the other group is required to satisfy a theoretical assumption imposed by decomposition methods referred to as the ``common support'' or ``overlapping support'' assumption. In this paper, I build on the existing literature to propose a generic decomposition method which relaxes the overlapping support assumption, allowing all observations to account for the actual observed gap. Furthermore, the decomposition framework that I present in this paper allows for decompositions in functionals of $Y$ -- e.g. permitting the study of how $X$ influences the gap in each quantile or portion of the distribution of $Y$. Although the set up is generic, my estimation strategy is focused on decomposing the gap in the distribution of $Y$, and I illustrate my method analysing the black-white wealth gap in the United States as a function of average lifetime labor income.

In order to estimate the full impact of lifetime labor income on differences in  wealth accumulation between blacks and whites, I make use of publicly available data from the Panel Study of Income Dynamics (PSID). Besides yearly household surveys dating from 1968, I also use the wealth supplement surveys from 1984, 1989 and 1994 to get information about the net wealth of American households, focusing on households where heads are either black or white with nation-wide representation. I follow the data pre-processing suggested by \citet{barskyboundcharleslupton2002} and I find that differences in the wealth distributions between blacks and whites are primarily due to differences in lifetime labor income, corroborating previous findings. However, by including observations outside of the common support of $X$, I find that: (i) the wealth distribution gap becomes bigger, especially for relatively higher values of wealth; and that (ii) observations outside of the common support contribute from $3\%$ to $19\%$ to the wealth gap distribution, non-trivial amounts that were previously ignored.

\textbf{Literature:} Decomposition methods in economics are still evolving, although little has been done with regards to relaxing the common support assumption (\citealp{fortinlemieuxfirpo2011}). \citet{barskyboundcharleslupton2002} are pioneers in thinking about this assumption. However, as shown in Section \ref{sec:ch3-methods}, functional decompositions similar to theirs (e.g. \citealp{dinardofortinlemieux1996}) assign virtually zero weight to observations outside the common support region, which is almost equivalent to trimming these observations. Furthermore, these methods still impose that the support for $X$ of one of the groups has to be a subset of/overlap the support of $X$ for the other group, which restricts the decomposition exercise, allowing it to be done using only white counterfactuals, but not black counterfactuals\footnote{The composition methods are set so the counterfactual quantity of group $W$, while having $B$ observable characteristics, is only identified when the covariates support of group $W$ is a subset of the covariate support for group $B$. For the black-white case, the support of black households' labor income does not overlap the whites' support, limiting the construction of the black counterfactual wealth distribution while having whites labor income. The method I propose allows for both counterfactual quantities to be estimated, not requiring the deletion of observations. Section \ref{sec:ch3-methods} defines the counterfactual quantity more precisely.}. In comparison, besides fully accounting for observations outside the common support, my estimator allows the decomposition to be done with either group counterfactual.

Another approach was developed in a way to deal with lack of overlapping support \citep{nopo2008,garcianoposalardi2009,FogelModenesi2021wagegap}, however this methodology applies only to decompositions of differences in the mean of $Y$. I build on \citet{nopo2008}'s set up to extend their approach to decompose differences in functionals of $Y$. I also propose a feasible estimation strategy to decompose differences between intergroup distributions of $Y$ leveraging established results for estimating conditional distributions developed by \citet{chernozhukovfernandezvalmelly2013} and implemented by \citet{chenchernozhukovfernandezvalmelly2016}. Although \citet{chernozhukovfernandezvalmelly2013} use their counterfactual distributional approach to decompose gaps as well, they also impose the common support assumption. This assumption is equally imposed by a relatively recent functional decomposition method using recentered influence functions \citet{firpofortinlemieux2018}.

Finally, in terms of application, I use the same dataset and data pre-processing decisions as \citet{barskyboundcharleslupton2002}, with the difference that, in addition to labor income, I also control for other observable covariates of the head of the household.

\textbf{Roadmap:} This paper is organized as follows. Section \ref{sec:ch3-methods} is the main section, setting up the framework that allows for decomposition in functionals of $Y$ while relaxing the common support assumption. This Section also discusses how other methods handle observations outside the common support and it proposes an estimator for decompositions in the distribution of $Y$. Next, Section \ref{sec:ch3-data} presents the PSID dataset used in this study, as well as details our sample cuts and descriptive statistics of the final sample. Section \ref{sec:ch3-results} presents results of my decomposition applied to the black-wealth distributional wealth gap, contrasting results using my estimator with the status quo. Finally, Section \ref{sec:ch3-conclusion} presents final thoughts and potential next steps in this research area.

%%%%%%%%%%%%%%%%%%%%%%%%%%%%%%%%%%%%%%%%%%%%%%%%%%%%%%%%%%%%%%%%%%%%%
\section{Methodology}\label{sec:ch3-methods}

A primitive form to study intergroup differences in a dependent variable of interest $Y$ can be done by studying differences in the distributions of $Y$ between groups. By computing counterfactual distributions and decomposing the intergroup gap in distributions, it is also possible to decompose a series of other gaps, namely gap in mean $Y$, or the gap in a certain quantile of $Y$. Therefore, in order to make my approach as flexible as possible, I choose to develop an identification and estimation strategy for decomposing differences in the distribution of $Y$.

%------------------------------------
\subsection{A distributional framework for decompositions}

A decomposition of difference in distributions allows the researcher to answer questions like ``what is the contribution of differences in $X$ (e.g. labor income) in the proportion of agents from each group at a certain level $y$ of $Y$ (e.g. at average wealth in the U.S., or for households at the top/bottom of the wealth distribution)?'' In order to perform this decomposition I first define my parameter of interest, i.e. the distributional gap on $Y$ between groups $W$ and $B$:
\begin{flalign}\label{eq:ch3-dist_gap}
\Delta(y) 		:=& H_W(y) - H_B(y) \nonumber \\
=& \int_{S_W} H_W(y|x) dF_W(x) - \int_{S_B} H_B(y|x) dF_B(x)
\end{flalign}
where $H_g(y)$ and $H_g(y|x)$ are the cumulative distribution function (c.d.f.) and the conditional c.d.f. of $Y$ for group $g \in \{B,W\}$; $F_g(x)$ is the c.d.f. of the group observable characteristics $X$ for group $g$; and $S_g$ is the support of $X$ for group $g$.

Decomposition methods typically decompose the gap into a portion due to differences in the \textit{composition} of $X$, $F_g$, fixing $H_g(y|x)$, and another portion due to \textit{structural} differences in the form of $H_g(y|x)$, while fixing the distribution of $X$ equal for both groups. In order compute these portions, I add and subtract the counterfactual distribution of $Y$, denoted by $H_0(y)$, to the gap function $\Delta(y)$ in equation \ref{eq:ch3-dist_gap}. This counterfactual quantity refers to the $Y$ distribution for members of group $W$, if they had group $B$'s distribution of $X$, more precisely defined as\footnote{Notice that the decomposition can also be performed by adding and subtracting the $B$ counterfactual distribution of $Y$, $\tilde H_0(y) := \int_{S_W} H_B(y|x) dF_W(x)$. This is the c.d.f. of $Y$ for the $B$ group, if group $B$ had group $W$ distribution of $X$.}:
\begin{equation}\label{eq:ch3-counterf-dist}
H_0(y) := \int_{S_B} H_W(y|x) dF_B(x),
\end{equation}

which allows the decomposition to be rewritten as:
\begin{flalign}\label{eq:ch3-dist_decomp}
\Delta(y) 		:=& H_W(y) - H_B(y) \pm H_0(y) \nonumber \\
=& \left[ H_W(y) - H_0(y) \right] + \left[ H_0(y) - H_B(y)\right] \nonumber \\
=& \underset{\Delta_X (y) := \text{Composition}}{  \underbrace{\left[ \int_{S_W} H_W(y|x) dF_W(x) - \int_{S_B} H_W(y|x) dF_B(x)  \right]} }  + \underset{\Delta_0 (y) := \text{Structure}}{  \underbrace{\left[ \int_{S_B} \left[H_A(y|x) -  H_B(y|x)\right]dF_B(x) \right]} }
\end{flalign}

In the wealth gap example, the composition portion, $\Delta_X (y)$, corresponds to the change in the proportion of whites with wealth equal to or below $y$, if whites had blacks' distribution of labor income. On the other hand, the structural portion, $\Delta_0 (y)$, can be interpreted as the difference in the proportions of blacks with wealth equal to or below $y$ if the wealth accumulation process for blacks shifted to how whites accumulate wealth conditional on income.

%------------------------------------
\subsection{Overlapping support assumption}

%..............
\subsubsection{Definition and purpose}
The overlapping support assumption, sometimes referred to as the common support assumption, guarantees that only comparable individuals -- in terms of $X$ -- from group $B$ are used to build the $W$ counterfactual distribution $H_0(y)$. More formally, considering the decomposition set up above, this assumption can be stated as: \\

\textbf{Definition}: \textit{Overlapping Support} (OS) \textit{Assumption} \\
Denote by $S_g$ the support of the distribution of $X$ for group $g \in \{B,W\}$, the supports of $X$ overlap if $S_W \supset S_B$.\footnote{Alternatively, if the decomposition is performed using group $B$ counterfactual distribution of $Y$, $\tilde H_0(y) := \int_{S_W} H_B(y|x) dF_W(x)$, then the overlapping support assumption becomes $S_W \subset S_B$.} \\

This allows the counterfactual $Y$ distribution $H_0(y)$, in equation \ref{eq:ch3-counterf-dist}, to be well defined as a c.d.f.. With $S_W$ being ``wider'' than than $S_B$, the integral in the counterfactual goes over regions of $x$ where the conditional distribution $H_W(y|x)$ also exists with positive measure, not allowing the integral over $S_B$ to go beyond 1.

%..............
\subsubsection{How does the literature handle it?}
This is a widely assumed assumption among dozens of decomposition methods, with the only exception of \citet{nopo2008} which only performs decomposition in mean $Y$. Little attention has been paid to this assumption, although the decomposition literature in economics is still active, especially when it comes to decompositions in functionals of $Y$ \citep{fortinlemieuxfirpo2011,firpofortinlemieux2018}.

Most of the existing decomposition methods handle failure of this assumption by simply trimming observations that do not lie in a region of \textit{common support} of $X$, defined by $S_B \cap S_W$ (e.g. \citealp{oaxaca1973}, \citealp{blinder1973}, \citealp{MachadoMata2005}, \citealp{chernozhukovfernandezvalmelly2013}, \citet{firpofortinlemieux2018}). In certain scenarios, trimming observations outside of the common support is reasonable, as it removes extreme outliers and it disregards a nearly negligible amount of observations, which would not change results significantly. However, in situations like the black-white wealth gap, where approximately $10\%$ of the observations lie outside of this region, simply deleting them can change results significantly. In other situations where unbiased counterfactuals are needed, researchers might need to add a large number of controls in order to mitigate potential violations of the assumption of \textit{selection on observables / ignorability}. This increases the dimensionality of $X$ and the complexity of the common support region, making it more likely that a bigger fraction of observations would assume $X$ values outside the common support (see \citet{FogelModenesi2021wagegap} for a more detailed discussion).

An alternative to deleting observations outside of the common support can be done using the estimation strategy developed by \citet{dinardofortinlemieux1996}, as pointed out by \citet{barskyboundcharleslupton2002}. This consists of a re-weighting approach, which rewrites the counterfactual distribution in \ref{eq:ch3-counterf-dist} as:
\begin{equation}\label{eq:ch3-dfl-counterf-dist}
H_0(y) = \int_{S_W} H_W(y|x) \underset{\psi(x)}{\underbrace{\frac{dF_B(x)}{dF_W(x)}}}dF_W(x),
\end{equation}

where $\psi(x)$ is the reweighting factor\footnote{The weight $\psi(x)$ can be interpreted as the Radon-Nikodym derivative of $dF_B$ with respect to $dF_W$, changing the measure of integration.} used to shift the distribution of $W$ to a counterfactual distribution, allowing the integral to be taken with respect to the measure of $W$, over the support of $W$, $S_W$. The weight can defined and rewritten using Bayes rule as follows:
\begin{flalign}\label{eq:ch3-dfl-weight}
\psi(x) = \frac{P(x|B)}{P(x|W)} = \frac{P(B|x)/P(B)}{P(W|x)/P(W)}
\end{flalign}
The literature uses the right hand side of equation \ref{eq:ch3-dfl-weight} to consistently estimate $\psi(x)$. Consider now an observation of from group $W$, assuming value $x^*$ that lies outside the region of common support, i.e. $x^* \notin S_B \cap S_W$ -- which is equivalent to $x^* \notin S_B$. This implies that $P(B|x^*) = 0$, making $\psi(x^*) = 0$, which is equivalent to trimming/deleting these observations outside the common support, as in the previous methods mentioned. In the black-white wealth gap, this corresponds to assigning zero weight to all whites at the top of the distribution of lifetime labor income.

%------------------------------------
\subsection{Relaxing the overlapping support assumption}
I propose a generalization in gap decompositions in functionals of $Y$, which relaxes the overlapping support assumption, allowing all observations to account for the actual observed gap in the decomposition exercise. The key insight to allow mismatching supports in the gap decomposition consists in rewriting the gap in equation \ref{eq:ch3-dist_decomp} having in mind 3 different portions of the support of $X$: (i) the common support region $S_W \cap S_B$, where observations in groups $B$ and $W$ can are comparable and can be used to construct counterfactual quantities of each other; (ii) the region outside the common support where only $W$ observations exist, $S_W \cap \bar S_B$; and (iii) the remaining region where only $B$ observations can be found $S_B \cap \bar S_W$, where $\bar S_g$ is the complement of $S_g$, $g \in \{ B,W\}$. I represent $H_g(y)$, $g \in \{ B,W\}$, as:
\begin{equation}\label{eq:ch3-cdf-support-regions}
H_g(y) = \int_{S_g \cap S_{\{j:j\neq g\}}} H_g(y | x) dF_g(x) + \int_{S_g \cap \bar S_{\{j:j\neq g\}}} H_g(y | x) dF_g(x) 
\end{equation}

This representation allows me to rewrite the right hand side of equation $\ref{eq:ch3-dist_decomp}$ in terms that are either in the common support, or out of it. The step by step identification of these terms is in the appendix \ref{app:sec:ch3-proof-identification-estimator}. The final result is the following generic decomposition:
\begin{flalign}\label{eq:ch3-detailed-generic-decomp}
\Delta(y) =: \Delta_X(y) + \Delta_0(y) + \Delta_W(y) + \Delta_B(y), 
\end{flalign}
where
\begin{flalign*}
&\Delta_X(y) 	= \int_{S_W \cap S_B} H_W(y | x) \left[\frac{dF_W(x)}{\mu_W(S_B)} - \frac{dF_B(x)}{\mu_B(S_W)}  \right] && \\
&\Delta_0(y) 	= \int_{S_W \cap S_B} \left[H_W(y | x) - H_B(y | x) \right] \frac{dF_B(x)}{\mu_B(S_W)} && \\
&\Delta_W(y) 	=  \left[ \int_{\bar S_B} H_W(y | x) \frac{dF_W(x)}{\mu_W(\bar S_B)} - \int_{S_B} H_W(y | x) \frac{dF_W(x)}{\mu_W(S_B)}  \right] \mu_W(\bar S_B) && \\
&\Delta_B(y) 	= \left[ \int_{S_W} H_B(y | x) \frac{dF_B(x)}{\mu_B(S_W)} - \int_{\bar S_W} H_B(y | x) \frac{dF_B(x)}{\mu_B(\bar S_W)}\right] \mu_B(\bar S_W) 	&& 	
\end{flalign*}
where $\mu_g(\cdot)$ is the probability of finding observations of group $g$ as a function of different parts of the support of $X$. It worth mentioning that if the supports of $X$ for both groups fully coincide, i.e. $S_B = S_W$, then my proposed set up collapses to the conventional set up in \ref{eq:ch3-dist_decomp}, used by decompositions in functionals of $Y$\footnote{Equation \ref{eq:ch3-detailed-generic-decomp} collapses to \ref{eq:ch3-dist_decomp}, if $S_B = S_W$, because in this situation, $\mu_B(\bar S_W)  = \mu_W(\bar S_B) =0$ and $\mu_W(S_B)=\mu_B(S_W) = 1$.}.

The interpretation of $\Delta_X(y)$ and $\Delta_0(y)$ in equation \ref{eq:ch3-detailed-generic-decomp} is exactly the same as in equation \ref{eq:ch3-dist_decomp}, only over the region of common support. The new terms, $\Delta_g(y)$, $g \in \{ B,W\}$ allow for observations outside of the common support to account for the existing gap. Precisely, $\Delta_g(y)$ corresponds to the change in the proportion of observations with $Y$ below or equal to the value $y$ if observations from group $g$ inside the common support suddenly get the $X$ distribution of observations from the same group outside the common support. In other words, this term captures differences between observations inside and outside of the common support region within the same group.

%------------------------------------
\subsection{Estimation strategy and plan for inference}

In this section I propose an econometric approach to consistently estimate each of the terms in the generic estimator and I lay out a plan for inference, to be followed in the next iteration of this project. 

In order to obtain the final estimator for the decomposition in this paper, I need to estimate two objects. Notice that all of the terms in the proposed estimator in equation \ref{eq:ch3-detailed-generic-decomp} are of the type 
\begin{equation}\label{eq:ch3-estimation-parameter}
\theta_g^{\tilde S}(y) := \int_{\tilde S} H_g(y | x) d\tilde F(x) = E_{\tilde F_x} [H_g(y | x)],
\end{equation}
where $\tilde S$ is an arbitrary region in the support of $X$ and $\tilde F(x)$ is a well defined c.d.f. over that region. This makes this quantity equivalent to an expectation of the conditional distribution function $H_g(y | x)$ with respect to the measure defined by $d \tilde F$, hence I need to estimate these two objects.

First, the econometric literature has developed greatly in terms of estimation and inference of conditional distribution functions such as $H_g(y | x)$. By employing distribution regressions it possible to estimate $H_g(y | x)$ consistently for each group separately \citep{chernozhukovfernandezvalmelly2013,chenchernozhukovfernandezvalmelly2016}\footnote{For further details in the estimation procedure for $H_g(y|x)$, see appendix \ref{app:sec:ch3-conditional-cdf-estimation}.}. Therefore the first step in my estimation procedure consist in fitting functional forms for $H_g(y | x)$, for each group.

For the second object, $d \tilde F(x)$, I can simply use the empirical c.d.f. to consistently estimate it. Take for example $\frac{dF_W(x)}{\mu_W(S_B)}$, which is equivalent to the measure of $x$ for a specific sub-population -- i.e. observation group $W$ that always lie in the common support. Assume that there are $N_W^{S_B}$ observations from this subpopulation, then each $x_i$ in this subpopulation is estimated to happen with empirical probability of $\frac{1}{N_W^{S_B}}$. Analogously, an estimator for $d \tilde F(x)$ is $\frac{1}{\tilde N}$, where $\tilde N$ corresponds to the number of observations of the subpopulation for which the measure $d \tilde F(x)$ is defined.

The two consistent estimators for $H_g(y | x)$ and for $d\tilde F(x)$ can be plugged into equation \ref{eq:ch3-estimation-parameter}, forming, by Slutsky theorem, a consistent estimator for $\theta_g^{\tilde S}(y)$:
\begin{equation}\label{eq:ch3-estimator-parameter}
\hat \theta_g^{\tilde S}(y) := \sum_{i=1}^{\tilde N} \frac{1}{\tilde N} \hat H_g(y | x_i)
\end{equation}

After obtaining $\hat \theta_g^{\tilde S}(y)$ for each group $g$ and different region of the support of $X$, the only remaining quantities to obtain estimates for all of the parameters in the gap decomposition in \ref{eq:ch3-detailed-generic-decomp} are $\mu_W(\bar S_B)$ and $\mu_B(\bar S_W)$, which can be consistently estimated as the proportions of observations of each group outside the common support. 	

As for inference, the most complex object in the estimation regards $\hat H_g(y | x_i)$ for which computing standard errors is not trivial. However, \citet{chernozhukovfernandezvalmelly2013} showed that the distribution regression coefficients are asymptotically jointly normally distributed and that the whole distribution regression process converges to a Gaussian process. This result can be useful to compute standard errors for each of the parameters in my proposed decomposition in the next iteration of this project.
% add an idea for inference

%------------------------------------
%\subsection{Defining common support}

%%%%%%%%%%%%%%%%%%%%%%%%%%%%%%%%%%%%%%%%%%%%%%%%%%%%%%%%%%%%%%%%%%%%%
\section{Data}\label{sec:ch3-data}

In order to illustrate the proposed decomposition method, I reassess the black-white wealth gap using the same data source and data preparation as described in \citet{barskyboundcharleslupton2002}. In particular, I make use of the publicly available yearly datasets from the Panel Study of Income Dynamics (PSID)\footnote{The public version of the yearly PSID datasets can be downloaded at \url{https://psidonline.isr.umich.edu/}.}. Since 1968, the PSID follows a nationally representative set of households in the United States, starting with nearly 5,000 households and asking questions mainly about family composition and income. In the years of 1984, 1989 and 1994, families were additionally surveyed about all assets and liability holdings that compose their net wealth, which is central information to this paper.

Although detailed wealth information was only available in the years of 1984, 1989 and 1994, I followed each household yearly since the beginning of the PSID, 1968, in order to obtain information regarding their labor income throughout their existence in the survey. This allowed me to calculate a proxy for lifetime labor earnings, by averaging each household's labor income (in prices of 1989), a quantity crucial to measure the impacts of labor income in the accumulated household wealth throughout decades. In addition to the wealth and labor income data, I also used the gender of the head of the household, education dummies and age as factors in the wealth decomposition -- contrasting with \citet{barskyboundcharleslupton2002}, who only used labor income as an independent variable in their exercise.

In order to compute the total net wealth of a household, several variables are summed: net business worth; personal bank accounts balances (including certificates of deposit, treasury bills, savings bonds); real estate equity value (deducted by mortgage and loans taken); value of stocks, investment funds, mutual funds, retirement accounts, of vehicles; and any other current debts.

Following \citet{barskyboundcharleslupton2002}, the sample is composed of households whose head is either black or white, from age 45 to 49, being head for at least 5 years prior each of the wealth survey years -- 1984, 1989 and 1994. The minimum of a 5 year window allows for more observations in order to estimate more precisely a proxy for lifetime labor income. The narrow age range restriction, on the other hand, has a few purposes. First, studying wealth differences as a function of labor income requires time for households to accumulate wealth, hence the relatively older age. Second, heads are significantly more likely to receive inheritances past the age of 50, which can be a confounding factor in this study, hence the upper threshold. Third, blacks were more likely to retire earlier, which can be another confounding factor that the age range aims to avoid. Fourth, the narrow age restriction is an attempt to control for other confounding factors related to age.

The remaining sample after all restrictions imposed consists in 382 households with black heads and 856 with white heads, as shown in table \ref{tab:ch3-demog_stats}. By construction, households from each race have similar age distribution, but they differ significantly with regards to the gender and education of their heads. Nearly half of the black households have female heads, while males are heads of $85.8\%$ of white households. Furthermore, $32\%$ of white heads have a college degree, in contrasting with only $7.6\%$ of black heads having one.

\begin{table}[htbp!]
	\centering
	\caption{Household head demographics by race: average age, proportion of females and proportion of college degree (standard deviation in parenthesis)}
	\begin{tabular}{crrr}
		\toprule
		& \multicolumn{1}{c}{\textbf{\quad Age}} & \multicolumn{1}{c}{\textbf{Female}} & \multicolumn{1}{c}{\textbf{College}} \\ \hline
		\textbf{Black} & 46.9                             & 0.485                               & 0.076                                \\
		{\scriptsize\textbf{N = 382}} & {\scriptsize \text{(1.4)}}                     & {\scriptsize\text{(0.500)}}                      & {\scriptsize\text{(0.265)}}                       \\ \midrule
		\textbf{White} & 46.9                             & 0.142                               & 0.320                                \\ 
		{\scriptsize\textbf{N = 856}} & {\scriptsize\text{(1.4)}}                     & {\scriptsize\text{(0.350)}}                      & {\scriptsize\text{(0.467)}}                       \\ \bottomrule
	\end{tabular}\label{tab:ch3-demog_stats}
	\flushleft \footnotesize \emph{Notes:} This table was calculated using the PSID. The sample comprises black and white household heads in 1984, 1989, or 1994, age 45-49. Households were also required to have at least five years of observations before each year. PSID sample weights are used in all comparisons.
\end{table}

For total net wealth and average labor income, their values have been converted to 1989 constant dollars using the Consumer Price Index for all Urban consumers (CPI-U). Mean average labor earnings for black households is $\$21,800$, corresponding to $56\%$ of white's average earnings, as shown by table \ref{tab:ch3-descriptive_stats}. As pointed out by \citet{barskyboundcharleslupton2002}, part of this discrepancy is due to black households having much lower marital rates, reducing their total labor income, relative to white households. The net wealth gap, on the other hand, is substantially bigger than the labor income gap between races. While whites have $\$208,600$ mean net wealth, black households have their mean wealth at only $17\%$ of white's wealth. Furthermore, the ratio of black-white labor income gap increases monotonically from $.01$ to $.76$ as we move from the 5th to the 95th percentiles of its distribution. Although the labor income ratio increases drastically by moving along the labor income distribution, the wealth ratio remains around $.20$ for the upper half of the wealth distribution for these groups. 

The aforementioned descriptive statistics raise a few questions that we hope to address in the following section. First, how can different portions of the wealth gap be decomposed into a portion due to labor income and other demographics, and another residual portion? For example, would the labor income be more important to explain the wealth gap at the bottom of the wealth distribution, versus at the top? Second, how to decompose the entire observed wealth gap, without trimming observations that do not share common support with regards to observable characteristics? For instance, instead of dropping blacks at the bottom of the labor earnings distribution and whites from the top, where the labor earnings support is not matched by the other race, how can we incorporate these observations to the wealth gap decomposition, precisely measuring their contribution to different parts of the gap distribution? 

\begin{table}[htbp!]
	\centering
	\caption{Sample Distribution Properties (\textit{in thousands of 1989 dollars})}
	\begin{tabular}{crrrrrr}
		\toprule
		\textbf{}            & \multicolumn{3}{c}{\textbf{Total net worth}}     & \multicolumn{3}{c}{\textbf{Avg labor earnings}} \\ \cline{2-4} \cline{5-7}
		& \textit{Black} & \textit{White} & \textit{Ratio} & \textit{Black}  & \textit{White}  & \textit{Ratio}  \\ \hline
		\textbf{Mean}        & 35.1           & 208.6          & .17           & 21.8            & 38.9            & .56            \\ \hline
		\textbf{Percentiles} &                &                &                &                 &                 &                 \\
		5th                  & -1.7           & 0.0              & .              & 0.1             & 10.2            & .01            \\
		10th                 & 0.0              & 4.0              & .00           & 3.4             & 15.4            & .22            \\
		25th                 & 1.3            & 33.5           & .04           & 9.7             & 25.0              & .39            \\
		50th                 & 19.2           & 92.0             & .21           & 17.1            & 37.0              & .46            \\
		75th                 & 38.1           & 193.3          & .20           & 33.8            & 49.5            & .68            \\
		90th                 & 84.5           & 402.7          & .21           & 44.2            & 61.6            & .72            \\
		95th                 & 120.5          & 638.0            & .19           & 55.0              & 72.2            & .76            \\ \hline
		\textbf{N =}         & 382            & 856.0            &    .45           & 382             & 856             &       .45          \\ \bottomrule
	\end{tabular}\label{tab:ch3-descriptive_stats}
	\flushleft \footnotesize \emph{Notes:} This table was calculated using the PSID and the CPI-U. The sample comprises black and white household heads in 1984, 1989, or 1994, age 45-49. Households were also required to have at least five years of observations before each year. PSID sample weights are used in all comparisons.
\end{table}

%%%%%%%%%%%%%%%%%%%%%%%%%%%%%%%%%%%%%%%%%%%%%%%%%%%%%%%%%%%%%%%%%%%%%
\section{Results}\label{sec:ch3-results}

%------------------------------------
\subsection{Common covariate support}

\begin{figure}[htbp!]
	\centering
	\caption{Probability distribution of Average Lifetime Labor Earnings}
	\includegraphics[width=.8\linewidth]{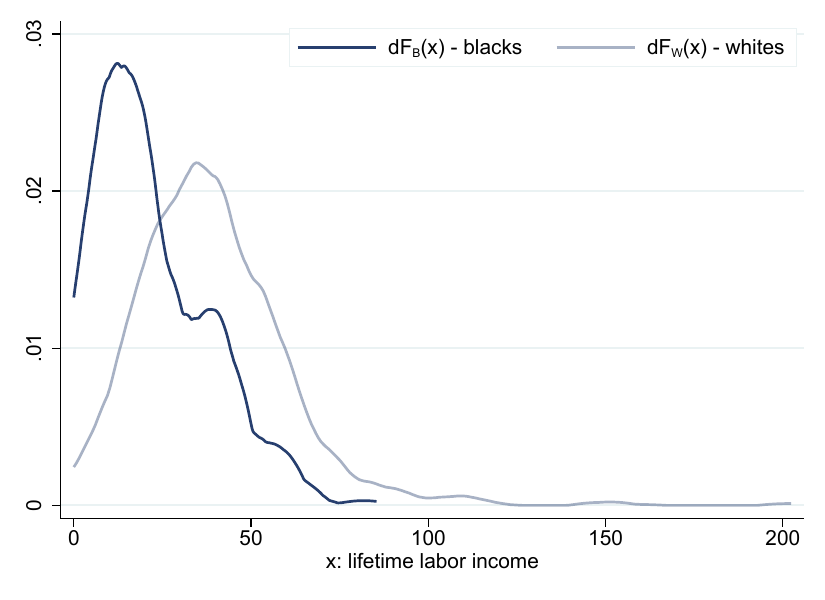}
	\label{fig:ch3-earnings_densities}
	\footnotesize\flushleft \emph{Notes:} PSID sample weights were used. Units correspond to thousands of 1989 dollars, computed using the CPI-U.
\end{figure}

As shown in table \ref{tab:ch3-descriptive_stats} and graphically illustrated in figure \ref{fig:ch3-earnings_densities}, the average lifetime labor income supports for black and whites do not fully overlap. There is an absence of black households at the top of the lifetime income distribution, as well as there are no whites at the very bottom of this distribution. Unmatched white households at the top of the lifetime labor income distribution correspond to $6.8\%$ of the total sample, while unmatched black households at the bottom represent $2.7\%$ of the total. Conventional decomposition methods would either trim or assign zero weight to these observations, resulting in not utilizing a total of $9.5\%$ of the sample.

%------------------------------------
\subsection{Gap decomposition}

\begin{figure}[htbp!]
	\centering
	\caption{Net wealth distribution for each race and counterfactual Net wealth distribution for white households}
	\includegraphics[width=.8\linewidth]{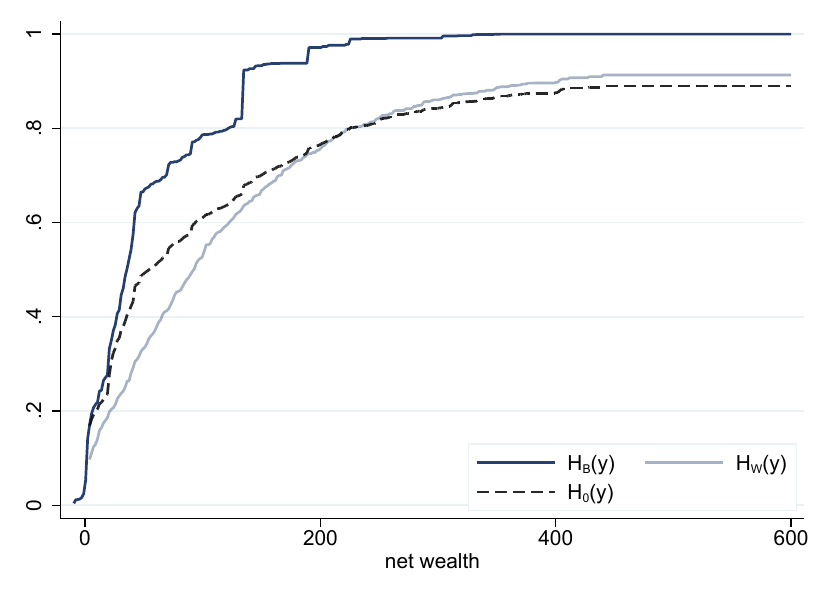}
	\label{fig:ch3-wealth_cdfs}
	\footnotesize\flushleft \emph{Notes:} PSID sample weights were used. Units correspond to thousands of 1989 dollars, computed using the CPI-U. The graph displays net wealth up to $\$600k$ for visual ease, encompassing approximately $97\%$ of the total sample. 
\end{figure}

The main ingredients of our parameter of interest $\Delta(y)$ are the wealth distribution functions $H_B(y)$, $H_W(y)$ and $H_0(y)$. We depict these functions for the black-white wealth gap decomposition in figure \ref{fig:ch3-wealth_cdfs}. This figure conveys a similar message conveyed by table \ref{tab:ch3-descriptive_stats}, that the wealth distribution of whites 1st-order stochastically dominates the wealth distribution for blacks, i.e. $H_W(y) \leq H_B(y), \forall y$. In other words, a bigger fraction of blacks are less wealthy than whites at any net wealth level, with the difference between these distributions reaching its biggest values around $\$50$ thousand dollars of 1989, and not vanishing completely even at high levels of wealth. Another important function depicted in figure \ref{fig:ch3-wealth_cdfs} is the estimate for $H_0(y)$, corresponding to the counterfactual wealth distribution of whites households, if they had the average labor income distribution of black households. It is noticeable that the counterfactual white wealth distribution is relatively close to the blacks' wealth distribution up to their medians, however, counterfactual white households at the top of the wealth distribution are still wealthier than blacks, despite their income being similar. % This result suggests that the composition effect $\Delta_X(y)$, which accounts for changes in wealth as consequence of imposing the same earnings distribution for both races, is stronger at the bottom of the wealth distribution, and that the structural effect $\Delta_0(y)$ does not vanish as we move towards the top of this distribution.

\begin{figure}[htbp!]
	\centering
	\caption{Net Wealth Distributional Gap relaxing versus imposing the overlapping support (OS) assumption}
	\includegraphics[width=.8\linewidth]{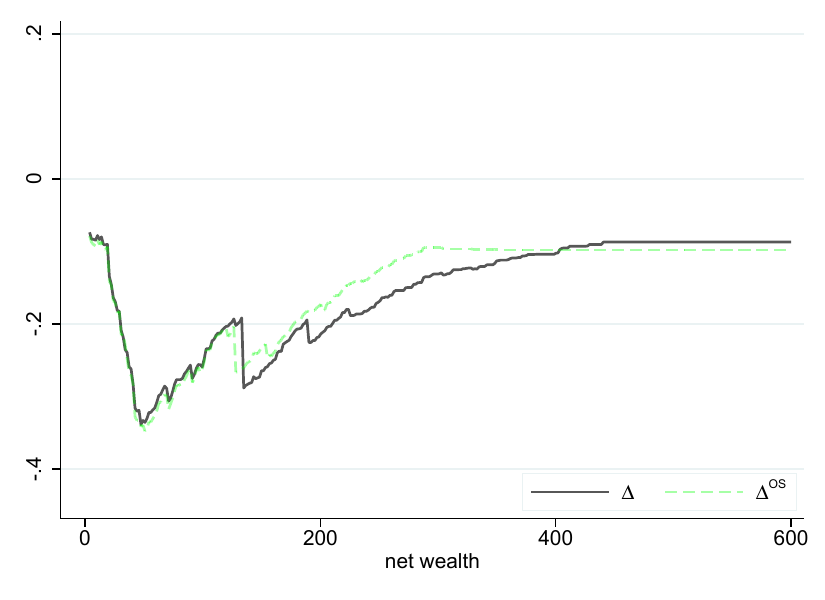}
	\label{fig:ch3-gaps_supcom}
	\footnotesize\flushleft \emph{Notes:} PSID sample weights were used. Units correspond to thousands of 1989 dollars, computed using the CPI-U. The graph displays net wealth up to $\$600k$ for visual ease, encompassing approximately $97\%$ of the total sample.
\end{figure}

Two versions of the gap in wealth distributions are plotted in figure \ref{fig:ch3-gaps_supcom}: one using all of the sample, denoted by $\Delta(y)$; and another deleting the observations outside of the common support, as done by traditional methods, denoted by $\Delta^{OS}(y)$. The solid black line is simply the difference between the c.d.f.s in figure \ref{fig:ch3-wealth_cdfs}, conveying the same message that the gap is more accentuated on relatively lower portions of the wealth support, and that the gap persists even for higher wealth values. In comparison, by imposing the Overlapping Support (OS) assumption, the gap shrinks towards zero for wealth levels between $\$150k$ and $\$400k$ (see green dashed line). If one believes in a positive correlation between average labor income and wealth, the attenuation in the gap is expected, since black households with the lowest average labor income and white households with the highest average labor income were not accounted by $\Delta^{OS}(y)$, mechanically shrinking this gap. Furthermore, with this figure it is possible to know exactly where these deleted observations are located in the wealth distribution. 

\begin{figure}[htbp!]
	\centering
	\caption{Comparison of decompositions: Net Wealth Distributional Gap Decomposition relaxing versus imposing the overlapping support (OS)}
	\includegraphics[width=.8\linewidth]{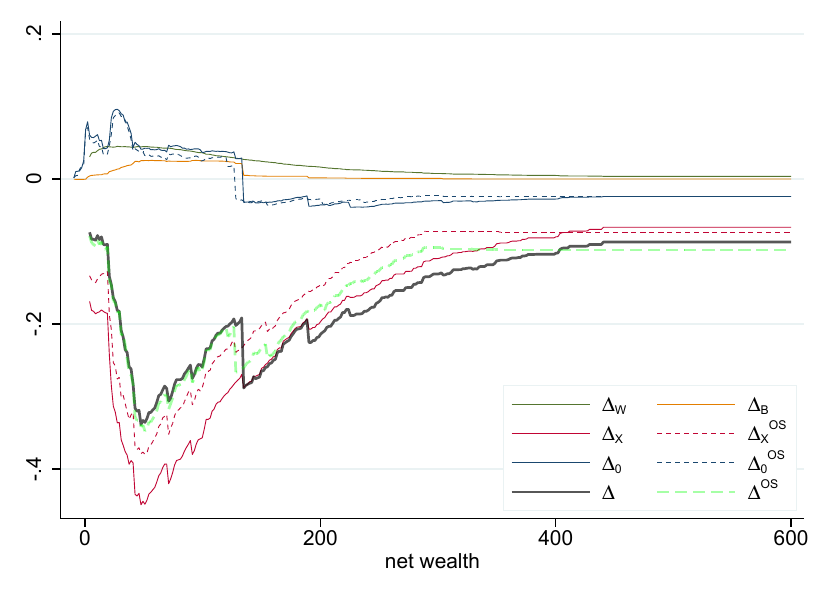}
	\label{fig:ch3-gap_decomp_comparison}
	\footnotesize\flushleft \emph{Notes:} PSID sample weights were used. Units correspond to thousands of 1989 dollars, computed using the CPI-U. The graph displays net wealth up to $\$600k$ for visual ease, encompassing approximately $97\%$ of the total sample.
\end{figure}

%\begin{figure}[htbp!]
%	\centering
%	\caption{Net Wealth Distributional Gap Decomposition relaxing the overlapping support assumption}
%	\includegraphics[width=.8\linewidth]{Chapter3/results/gap_decomp_allcov1_counterfmode2}
%	\label{fig:ch3-gap_decomp_mine}
%	\footnotesize\flushleft \emph{Notes:} PSID sample weights were used. Units correspond to thousands of 1989 dollars, computed using the CPI-U.
%\end{figure}
Now I analyze results regarding the decomposition of the gaps $\Delta(y)$ and $\Delta^{OS}(y)$. I add the decomposition of these two terms into composition portions ($\Delta_X(y)$ and $\Delta_X^{OS}(y)$) and structural portions ($\Delta_0(y)$ and $\Delta_0^{OS}(y)$) in figure \ref{fig:ch3-gap_decomp_comparison}. For the decomposition of the gap relaxing the overlapping support assumption, I also compute and plot the contribution of the observations outside the common support, denoted by $\Delta_B(y)$ and $\Delta_W(y)$, in figure \ref{fig:ch3-gap_decomp_comparison}. The terms regarding the new decomposition method were drawn with solid lines, and the terms representing the conventional decomposition methods imposing the OS assumption were drawn using dashed lines. 

Among all the decomposition terms defined in equation \ref{eq:ch3-detailed-generic-decomp}, the major contributor to the distributional wealth gap is by far the composition effect, either imposing the OS assumption or not, as denoted by the red lines in figure \ref{fig:ch3-gap_decomp_comparison}. This effect peaks at around $\$50k$ and reduces as wealth values increase, but never vanishes. In practical terms, white households would be substantially less wealthy if they earned as black households do from their jobs throughout their lifetime. By relaxing the OS assumption, the composition effect of $X$ is magnified due to including more extreme observations with regards to their average labor income.

The structural effects are represented by blue lines in figure \ref{fig:ch3-gap_decomp_comparison} and their magnitudes do not change much whether imposing OS or not. These effects capture differences in how white households versus black households accumulate wealth for similar values of labor income, i.e. $H_W(y|x) - H_B(y|x)$, weighted by blacks average labor income distribution. The fact that the blue curves are similar indicates that deleting the extreme observations outside the common support does not alter significantly how these conditional distribution functions are fitted. In terms of the shape of $\Delta_0(y)$ and $\Delta_0^{OS}(y)$, they both indicate that if households accumulate wealth like whites, while earning like blacks, they would be poorer than black households until approximately $\$180k$ of wealth. After this point, i.e. for relatively wealthier households, accumulating wealth like white households makes them wealthier.

Finally I discuss the terms that indicate the direct impact of observations outside the common support into the distributional wealth gap, plotted with the solid green and yellow curves in figure \ref{fig:ch3-gap_decomp_comparison}. Both of these $\Delta_B(y)$ and $\Delta_W(y)$ effects act in opposite directions as the observed gap, being positive, especially before $\$180k$ of wealth, approximating zero as wealth values increase. Due to a relatively larger portion of whites being outside of the common support area, with high labor income, the contribution of this group of observations is bigger in magnitude than the contribution of blacks outside of the common support. $\Delta_W(y)$ being positive indicates that for a given level $y$ of wealth, whites outside the common support accumulate wealth at a slower rate than whites within the common support. Alternatively, $\Delta_B(y)$ greater than zero indicates that blacks in the common support accumulate wealth at a slower rate than blacks outside the common support.

\begin{figure}[htbp!]
	\centering
	\caption{Contribution of each factor to the Net Wealth Distributional Gap Decomposition}
	\includegraphics[width=.8\linewidth]{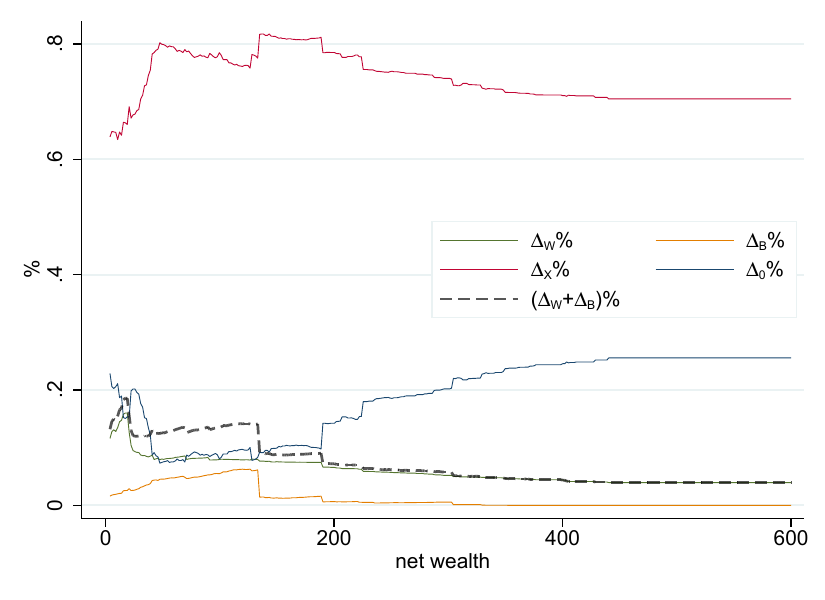}
	\label{fig:ch3-gap_decomp_contributions}
	\footnotesize\flushleft \emph{Notes:} PSID sample weights were used. Units correspond to thousands of 1989 dollars, computed using the CPI-U. The contribution of each factor was computed as the ratio of the absolute value of that factor, divided by the sum of the absolute value of all other factors. The graph displays net wealth up to $\$600k$ for visual ease, encompassing approximately $97\%$ of the total sample.
\end{figure}

The participation in percentage terms of each of the decomposition factors $\Delta_X(y)$, $\Delta_0(y)$, $\Delta_B(y)$ and $\Delta_W(y)$ into the distributional gap $\Delta(y)$ for the new estimator is plotted in figure \ref{fig:ch3-gap_decomp_contributions}. The contribution of each factor was computed as the fraction between the absolute value of the factor and the sum of the absolute values of all the factors, capturing the strength at which each factor shifts the gap. As expected, differences in the composition of the labor income $X$ accounts on average for approximately $75\%$ of the wealth gap (red solid line). Except for the composition effect, the contribution of the structural portion prevails among others for wealth values beyond $\$200k$ (blue solid line), ranging from $15\%$ to $26\%$ after this wealth value. After the composition effect, however, the contribution of the terms regarding the lack of common support, i.e. $\Delta_B(y) + \Delta_W(y)$, tend to be the second most important effect for wealth values until $\$175k$, where $75\%$ for the sample is located (black dashed line). This indicates the importance of including these observations in the gap decomposition analysis. Considering all wealth values, the contribution of these new terms ranges from $3\%$ to $19\%$ of the overall gap.

%%%%%%%%%%%%%%%%%%%%%%%%%%%%%%%%%%%%%%%%%%%%%%%%%%%%%%%%%%%%%%%%%%%%%
\section{Conclusion}\label{sec:ch3-conclusion}

This paper expands the set of intergroup gap decomposition tools in Economics by proposing a new distributional decomposition method that relaxes the Overlapping Support (OS) assumption. Current decomposition methods impose the OS assumption, which results in deleting or assigning zero weight to observations from one group unmatched by observations in the other group in terms of observable characteristics. These observations, however, constitute part of the observed gap, hence deleting them changes the gap itself, as well as modifies each of the terms in the decomposition. On the other hand, the new decomposition estimator that I propose allows for observations outside the common covariate support to participate in the decomposition of the observed gap. The key insight to obtain this estimator regards rewriting the integral inside the gap, with respect to covariates , while splitting the covariates support into regions where groups share common support and where supports do not match. I show that the estimation procedure of each portion of the decomposition requires the estimation of (i) distributions of covariates $X$ over different regions of the support, and of (ii) the conditional c.d.f. of the variable of interest $Y$ given $X$. I estimate (i) using analogous empirical c.d.f. and (ii) using distribution regressions.

I illustrate the proposed method studying the black-white wealth gap in the United States, using the publicly available income and wealth data from the Panel Study of Income Dynamics (PSID) from 1968 to 1994. More specifically, I assess how differences in lifetime labor income between races contribute to the observed difference in their wealth accumulation, while also accounting for black and white households that have extreme values of labor income. I show that households with extreme values of income represent nearly $10\%$ of the sample, and current decomposition methods would disregard these observations, attenuating the existing gap and changing the portions it is decomposed to. Moreover, I show that, although intergroup differences in average labor income predominantly explains the wealth gap, the terms associated with the pure effect of observations outside of the common support in the decomposition contributes from $3\%$ to $19\%$ to the overall gap, reaching its highest values in parts of the wealth distribution where most of the sample is concentrated.

% ======================================================================
% References
% ======================================================================
%\bibliographystyle{apalike}
%\bibliography{references}
%\bibliography{/home/bm/Dropbox\ \(University\ of\ Michigan\)/Kaizen/references.bib}

\clearpage
\bibliographystyle{aer}	
\bibliography{References.bib}

@article{MachadoMata2005,
	title        = {Counterfactual decomposition of changes in wage distributions using quantile regression},
	author       = {Machado, José A. F. and Mata, José},
	year         = 2005,
	journal      = {Journal of Applied Econometrics},
	volume       = 20,
	number       = 4,
	pages        = {445--465},
	doi          = {https://doi.org/10.1002/jae.788},
	url          = {https://onlinelibrary.wiley.com/doi/abs/10.1002/jae.788},
	eprint       = {https://onlinelibrary.wiley.com/doi/pdf/10.1002/jae.788}
}

@misc{FogelModenesi2021wagegap,
      title={Detailed Gender Wage Gap Decompositions: Controlling for Worker Unobserved Heterogeneity Using Network Theory}, 
      author={Jamie Fogel and Bernardo Modenesi},
      year={2024},
      eprint={2405.04365},
      archivePrefix={arXiv},
      primaryClass={econ.EM}
}

@techreport{garcianoposalardi2009,
	title        = {{Gender and Racial Wage Gaps in Brazil 1996-2006: Evidence Using a Matching Comparisons Approach}},
	author       = {Luana Marquez Garcia and Hugo \~{N}opo and Paola Salardi},
	year         = 2009,
	month        = May,
	number       = 4626,
	doi          = {},
	url          = {https://ideas.repec.org/p/idb/wpaper/4626.html},
	institution  = {Inter-American Development Bank, Research Department},
	type         = {Research Department Publications}
}

@misc{chenchernozhukovfernandezvalmelly2016,
	title        = {Counterfactual: An R Package for Counterfactual Analysis},
	author       = {Mingli Chen and Victor Chernozhukov and Iván Fernández-Val and Blaise Melly},
	year         = 2016,
	eprint       = {1610.07894},
	archiveprefix = {arXiv},
	primaryclass = {stat.CO}
}

@article{barskyboundcharleslupton2002,
	title        = {Accounting for the Black-White Wealth Gap: A Nonparametric Approach},
	author       = {Robert Barsky and John Bound and Kerwin Kofi Charles and Joseph P. Lupton},
	year         = 2002,
	journal      = {Journal of the American Statistical Association},
	publisher    = {[American Statistical Association, Taylor & Francis, Ltd.]},
	volume       = 97,
	number       = 459,
	pages        = {663--673},
	issn         = {01621459},
	url          = {http://www.jstor.org/stable/3085702}
}

@article{firpofortinlemieux2018,
	title        = {{Decomposing Wage Distributions Using Recentered Influence Function Regressions}},
	author       = {Sergio P. Firpo and Nicole M. Fortin and Thomas Lemieux},
	year         = 2018,
	month        = {May},
	journal      = {Econometrics},
	volume       = 6,
	number       = 2,
	pages        = {1--40},
	doi          = {},
	url          = {https://ideas.repec.org/a/gam/jecnmx/v6y2018i2p28-d149033.html}
}

@article{nopo2008,
	title        = {Matching as a Tool to Decompose Wage Gaps},
	author       = {Hugo {\~N}opo},
	year         = 2008,
	journal      = {The Review of Economics and Statistics},
	publisher    = {The MIT Press},
	volume       = 90,
	number       = 2,
	pages        = {290--299},
	issn         = {00346535, 15309142},
	url          = {http://www.jstor.org/stable/40043147}
}

@article{blinder1973,
	title        = {Wage Discrimination: Reduced Form and Structural Estimates},
	author       = {Alan S. Blinder},
	year         = 1973,
	journal      = {The Journal of Human Resources},
	publisher    = {[University of Wisconsin Press, Board of Regents of the University of Wisconsin System]},
	volume       = 8,
	number       = 4,
	pages        = {436--455},
	issn         = {0022166X},
	url          = {http://www.jstor.org/stable/144855}
}

@article{oaxaca1973,
	title        = {Male-Female Wage Differentials in Urban Labor Markets},
	author       = {Ronald Oaxaca},
	year         = 1973,
	journal      = {International Economic Review},
	publisher    = {[Economics Department of the University of Pennsylvania, Wiley, Institute of Social and Economic Research, Osaka University]},
	volume       = 14,
	number       = 3,
	pages        = {693--709},
	issn         = {00206598, 14682354},
	url          = {http://www.jstor.org/stable/2525981}
}

@incollection{fortinlemieuxfirpo2011,
	title        = {Chapter 1 - Decomposition Methods in Economics},
	author       = {Nicole Fortin and Thomas Lemieux and Sergio Firpo},
	year         = 2011,
	publisher    = {Elsevier},
	series       = {Handbook of Labor Economics},
	volume       = 4,
	pages        = {1--102},
	doi          = {https://doi.org/10.1016/S0169-7218(11)00407-2},
	issn         = {1573-4463},
	url          = {https://www.sciencedirect.com/science/article/pii/S0169721811004072},
	editor       = {Orley Ashenfelter and David Card}
}

@article{dinardofortinlemieux1996,
	title        = {Labor Market Institutions and the Distribution of Wages, 1973-1992: A Semiparametric Approach},
	author       = {John DiNardo and Nicole M. Fortin and Thomas Lemieux},
	year         = 1996,
	journal      = {Econometrica},
	publisher    = {[Wiley, Econometric Society]},
	volume       = 64,
	number       = 5,
	pages        = {1001--1044},
	issn         = {00129682, 14680262},
	url          = {http://www.jstor.org/stable/2171954}
}

@article{chernozhukovfernandezvalmelly2013,
	title        = {Inference on Counterfactual Distributions},
	author       = {Chernozhukov, Victor and Fernández-Val, Iván and Melly, Blaise},
	year         = 2013,
	journal      = {Econometrica},
	volume       = 81,
	number       = 6,
	pages        = {2205--2268},
	doi          = {https://doi.org/10.3982/ECTA10582},
	url          = {https://onlinelibrary.wiley.com/doi/abs/10.3982/ECTA10582},
	eprint       = {https://onlinelibrary.wiley.com/doi/pdf/10.3982/ECTA10582}
}

% ======================================================================
% ======================================================================
% ======================================================================
% ======================================================================
% ======================================================================
% ======================================================================

\clearpage
\appendix
\appendixpage

% ======================================================================
% ======================================================================

\section{Identification proof of the estimator}\label{app:sec:ch3-proof-identification-estimator}

This section develops an intergroup gap decomposition in functionals of $Y$, namely $H_g(y)$, $g \in \{ B,W\}$ that allows for observations outside of the common support to be accounted. The steps taken to develop the estimator are shown below:
\begin{flalign*}\label{app:eq:ch3-proof-identification-estimator}
	\Delta (y) 		:=&  H_W(y) - H_B(y) 											&& \\
	=&  \int_{S_W} H_W(y | x) dF_W(x) - \int_{S_B} H_B(y | x) dF_B(x) 	&& \\
	=&  \left[\int_{S_W \cap \bar S_B} H_W(y | x) dF_W(x) + \int_{S_W \cap S_B} H_W(y | x) dF_W(x)\right] && \\
	&\phantom{A+}- \left[\int_{\bar S_W \cap S_B} H_B(y | x) dF_B(x) + \int_{S_W \cap S_B} H_B(y | x) dF_B(x)\right] 	&& \\			
	=&  \int_{\bar S_B} H_W(y | x) dF_W(x) + \int_{S_W \cap S_B} \left[H_W(y | x) dF_W(x) - H_B(y | x) dF_B(x)\right] && \\
	& \quad - \int_{\bar S_W} H_B(y | x) dF_B(x) 	&& 				\\
	=&  \int_{\bar S_B} H_W(y | x) \frac{dF_W(x)}{\mu_W(\bar S_B)} \mu_W(\bar S_B) + && \\
	& \quad + \int_{S_W \cap S_B} \left[H_W(y | x) \frac{dF_W(x)}{\mu_W(S_B)} \mu_W(S_B) - H_B(y | x) \frac{dF_B(x)}{\mu_B(S_W)} \mu_B(S_W) \right] + && \\
	& \quad - \int_{\bar S_W} H_B(y | x) \frac{dF_B(x)}{\mu_B(\bar S_W)} \mu_B(\bar S_W) 	&& 				\\
	=&  \int_{\bar S_B} H_W(y | x) \frac{dF_W(x)}{\mu_W(\bar S_B)} \mu_W(\bar S_B) + \\
	& \quad + \int_{S_W \cap S_B} \left[H_W(y | x) \frac{dF_W(x)}{\mu_W(S_B)} \left[1 - \mu_W(\bar S_B)\right] - H_B(y | x) \frac{dF_B(x)}{\mu_B(S_W)} \left[1 - \mu_B(\bar S_W)\right] \right] + && \\
	& \qquad - \int_{\bar S_W} H_B(y | x) \frac{dF_B(x)}{\mu_B(\bar S_W)} \mu_B(\bar S_W) 	&& 				\\			
	=&  \left[ \int_{\bar S_B} H_W(y | x) \frac{dF_W(x)}{\mu_W(\bar S_B)} - \int_{S_W \cap S_B} H_W(y | x) \frac{dF_W(x)}{\mu_W(S_B)}  \right] \mu_W(\bar S_B) && \\
	& \qquad + \int_{S_W \cap S_B} \left[H_W(y | x) \frac{dF_W(x)}{\mu_W(S_B)} - H_B(y | x) \frac{dF_B(x)}{\mu_B(S_W)}  \right] && \\
	& \qquad + \left[ \int_{S_W \cap S_B} H_B(y | x) \frac{dF_B(x)}{\mu_B(S_W)} - \int_{\bar S_W} H_B(y | x) \frac{dF_B(x)}{\mu_B(\bar S_W)}\right] \mu_B(\bar S_W) 	&& 				\\		
	=&  \left[ \int_{\bar S_B} H_W(y | x) \frac{dF_W(x)}{\mu_W(\bar S_B)} - \int_{S_W \cap S_B} H_W(y | x) \frac{dF_W(x)}{\mu_W(S_B)}  \right] \mu_W(\bar S_B) && \\
	& \qquad + \int_{S_W \cap S_B} \left[H_W(y | x) \frac{dF_W(x)}{\mu_W(S_B)} - H_B(y | x) \frac{dF_B(x)}{\mu_B(S_W)}  \pm H_W(y | x) \frac{dF_B(x)}{\mu_B(S_W)} \right]  && \\
	& \qquad + \left[ \int_{S_W \cap S_B} H_B(y | x) \frac{dF_B(x)}{\mu_B(S_W)} - \int_{\bar S_W} H_B(y | x) \frac{dF_B(x)}{\mu_B(\bar S_W)}\right] \mu_B(\bar S_W) 	&& 				\\		
	=&  \underset{\Delta_W(y)}{\underbrace{\left[ \int_{\bar S_B} H_W(y | x) \frac{dF_W(x)}{\mu_W(\bar S_B)} - \int_{S_B} H_W(y | x) \frac{dF_W(x)}{\mu_W(S_B)}  \right] \mu_W(\bar S_B)}} && \\
	& \qquad + \underset{\Delta_X(y)}{\underbrace{\int_{S_W \cap S_B} H_W(y | x) \left[\frac{dF_W(x)}{\mu_W(S_B)} - \frac{dF_B(x)}{\mu_B(S_W)}  \right]}} && \\
	& \qquad + \underset{\Delta_0(y)}{\underbrace{\int_{S_W \cap S_B} \left[H_W(y | x) - H_B(y | x) \right] \frac{dF_B(x)}{\mu_B(S_W)} }} && \\
	& \qquad + \underset{\Delta_B(y)}{\underbrace{\left[ \int_{S_W} H_B(y | x) \frac{dF_B(x)}{\mu_B(S_W)} - \int_{\bar S_W} H_B(y | x) \frac{dF_B(x)}{\mu_B(\bar S_W)}\right] \mu_B(\bar S_W)}} 	&& 				\\				
	=& \Delta_W(y) + \Delta_X(y) + \Delta_0(y) + \Delta_B(y)	&&
\end{flalign*}

\begin{comment}
	Each term can be also described as:
	\begin{flalign}
		&\Delta_W(y) =  \left\{ E_B \left[H_W(y | x) | Unmatched_B \right] - E_B \left[H_W(y | x) | Matched_B \right]  \right\} P_B \left( Unmatched\right) && \\
		&\Delta_X(y) = E_B \left[H_W(y | x) | Matched_B \right] - E_A \left[H_W(y | x) | Matched_A \right]  && \\
		&\Delta_0(y) = E_A \left[H_W(y | x) | Matched_A \right] - E_A \left[H_B(y | x) | Matched_A \right]  && \\
		&\Delta_B(y) = \left\{ E_A \left[H_B(y | x) | Matched_A \right] - E_A \left[H_B(y | x) | Unmatched_A \right]  \right\} P_A \left( Unmatched \right) && 
	\end{flalign}
\end{comment}

\newpage

\section{Estimation of the conditional distribution function}\label{app:sec:ch3-conditional-cdf-estimation}

The estimation of $H_g(y|x)$ consists in imposing the flexible functional assumption $H_g(y|x) := \Lambda(T(x)' \alpha(y))$, where $\Lambda(\cdot)$ is a link function of choice, $\alpha(\cdot)$ is an unknown function-valued parameter and $T(\cdot)$ is a vector of transformations of $X$. This distribution regression model is quite flexible, since $H_g(y|x)$ can be approximated arbitrarily close enough depending on the richness of $T$, e.g. polynomials or B-splines of high degree. This makes the correct choice of $\Lambda$ less crucial, given a sufficiently rich $T$. 

The estimation of $\alpha(y)$, and consequently of $H_g(y|x)$, can be done with a simple maximum likelihood approach. The likelihood with respect to $\alpha(y)$ can be written as:
\begin{flalign*}
	\mathcal{L}(Y,X|\alpha(y)) = \Pi_{i \in A} P(Y_i \leq y | X_i, \alpha(y))^{\mathbbm{1}\{Y_i \leq y\} } \left[1-P(Y_i > y | X_i, \alpha(y))\right]^{\mathbbm{1}\{Y_i > y\} }
\end{flalign*}

Replacing the probabilities in the likelihood with a known link function, and taking logs, then the estimator is the solution of:
\begin{flalign}
	\hat{\alpha}(y) = \arg\max_{\alpha \in \Re^{d_x}} \sum_{i \in A} \mathbbm{1}\{Y_i \leq y\} \Lambda(T(x_i)'\alpha) + \mathbbm{1}\{Y_i > y\} \left[1-\Lambda(T(x_i)'\alpha)\right]
\end{flalign}

The theory behind the estimation procedure is described by \citet{chernozhukovfernandezvalmelly2013} and I use the command \textit{drprocess} in Stata to implement it. Details of the implementation can be also found in \citet{chenchernozhukovfernandezvalmelly2016}.

\end{document}